# Neuronal control of locomotor handedness in Drosophila

Sean M. Buchanan[1], Jamey S. Kain[1] and Benjamin L. de Bivort[1,2] *

1 The Rowland Institute at Harvard, 100 Edwin Land Blvd., Cambridge, MA 02142

2 Center for Brain Science and Department of Organismic and Evolutionary Biology, Harvard University, 52 Oxford St, Cambridge, MA 02138

* correspondence: debivort@oeb.harvard.edu

**Abstract**

Handedness in humans – better performance using either the left or right hand – is personally familiar, moderately heritable[1], and regulated by many genes[2], including those involved in general body symmetry[3]. But behavioral handedness, i.e. lateralization, is a multifaceted phenomenon. For example, people display clockwise or counter-clockwise biases in their walking behavior that is uncorrelated to their hand dominance[4,5], and lateralized behavioral biases have been shown in species as disparate as mice (paw usage[6]), octopi (eye usage[7]), and tortoises (side rolled on during righting[8]). However, the mechanisms by which asymmetries are instilled in behavior are unknown, and a system for studying behavioral handedness in a genetically tractable model system is needed. Here we show that *Drosophila melanogaster* flies exhibit striking variability in their left-right choice behavior during locomotion. Very strongly biased "left-handed" and "right-handed" individuals are common in every line assayed. The handedness of an individual persists for its lifetime, but is not passed on to progeny, suggesting that mechanisms other than genetics determine individual handedness. We use the *Drosophila* transgenic toolkit to map a specific set of neurons within the central complex that regulates the strength of behavioral handedness within a line. These findings give insights into choice behaviors and laterality in a simple model organism, and demonstrate that individuals from isogenic populations reared under experimentally identical conditions nevertheless display idiosyncratic behaviors.

In order to investigate whether flies display individual left-right locomotor biases, we developed a simple, high throughput assay to quantify turning. Flies were placed individually in Y-shaped mazes, allowed to walk freely for two hours, with their centroids tracked in two dimensions (Fig 1a-c, Supplemental Movie 1). Each maze was symmetrical and evenly lit, so that choices were unbiased rather than stimulus-driven. The fraction of times the fly passed through the center of the maze and chose to go right defined a "turn bias score" (Fig 1d). Each fly typically performed hundreds of choices per experiment (Fig S1a). Precise quantification of the distribution of individual behaviors requires high sample sizes, so many mazes were arrayed in parallel (Fig S1b-d). Thus, our results reflect over 25,000 individual flies and 16,000,000 turn choices.

We measured the turn biases of hundreds of individual flies from seven different fly lines: Berlin-K (BK), Canton-S (CS), Cambridge-A (CA,[9]), two lines of CS that were independently inbred for 10 generations, CA that was inbred for 10 generations[10], and $w^{1118}$, the background line for many transgenic flies (Figs 1e-f, S1e). As expected, the average probability of turning right (the turning score), averaged across all individuals within each line was ~50%. However, this belied profound individual-to-individual variability, and an individual fly's probability of turning right often diverged markedly from the population average. For example, nearly one quarter (23.5%) of CS flies turned right greater than 70% of the time or less than 30% of the time. This is an unlikely outcome indeed if all flies were choosing to turn right with identical probabilities. This null hypothesis can be modeled using the binomial distribution, with each fly performing $n_i$ choices (equal to the number it performed in the experiment) and a probability of turning right $p$ (equal to the mean probability observed across all flies). This is statistically justified because sequential turns were essentially independent of one another (Fig S1f). Compared to this null hypothesis, biased "righty" and "lefty" individuals are vastly over-represented ($p < 10^{-16}$, $10^{-4}$ by $\chi^2$ test of variance and bootstrap resampling respectively).

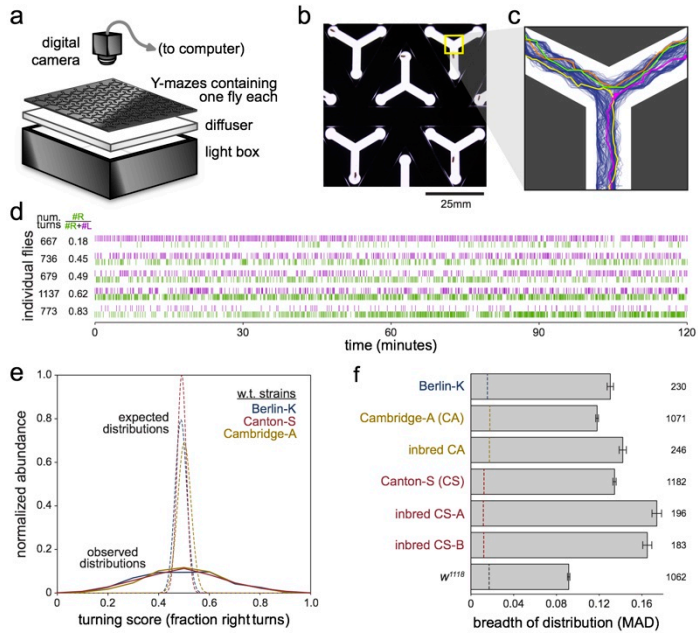
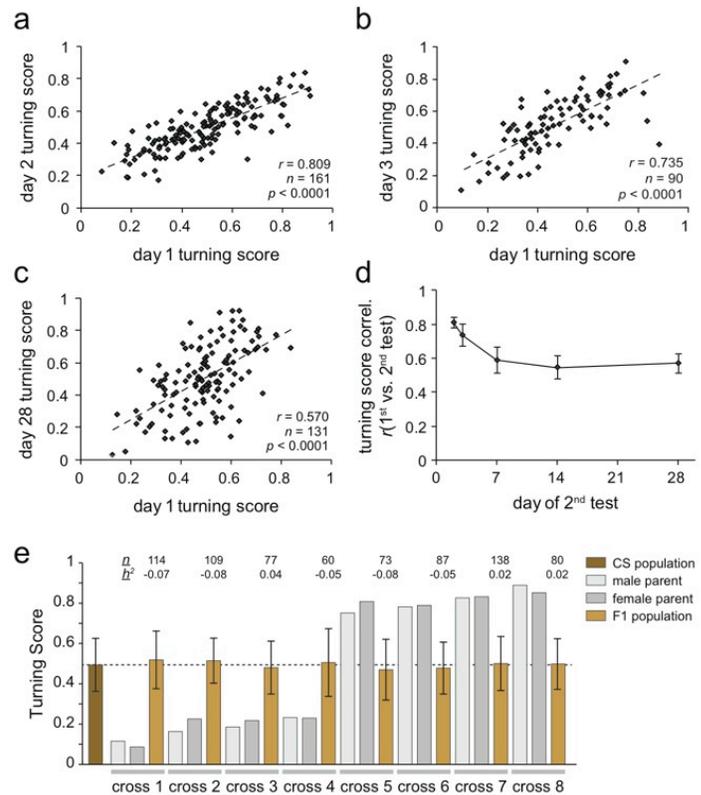

**Figure 1 – Individual flies exhibit biases in left-right turning. a**) Schematic of a device for assaying left-right turning in individuals. Flies were placed into an array containing many individual Y-mazes. The mazes were illuminated from below and imaged from above, and the positions of the flies were recorded. **b**) Detail of Y-mazes containing individual flies. **c**) 100 example turn paths through the Y-maze recorded from a single fly over 2 hours (blue). Other colors highlight individual turns. **d**) Left and right turn sequences for example flies of varying turn biases. **e**) Observed distribution of turning scores (fraction right turns) measured from wild type lines (solid lines), and corresponding expected distributions of turning scores (dashed lines). Sample sizes given in **f**. BK: Berlin-K, CA: Cambridge-A, CS: Canton-S. **f**) The breadth of the distribution of turning scores for seven lines as measured by mean absolute deviation from the mean (MAD). Error bars are +/- 1 standard error estimated by bootstrap resampling. Dashed lines indicate MADs expected under a binomial null model. All lines other than $w^{1118}$ (transgenic background line) are nominally wild type.

We were unable to identify any trivial sources of left-right turning bias. Neither the light boxes, nor the maze arrays, nor the positions of the mazes within the arrays had any effect on the observed mean turning bias (Figs S1b-d). Anosmic flies[11] displayed the same variability as control flies (Fig S1g), suggesting that flies were not following odor trails within the mazes. Lastly, general health or activity level did not explain the strong biases of flies; there is no correlation between turning score and number of turns completed in the 2hour experiment (Fig S1h).

Next, we evaluated the persistence of turning biases. Individual flies were tested in the Y-mazes, recovered, stored individually, and then tested a second time, in a different maze, either 1, 2, 6, 13 or 27 days later. Individual turning scores were highly correlated across time, and range from $r=0.57$ for day 1 vs. day 28 to $r=0.81$ for day 1 vs. day 2 (all $p<0.0001$, Fig 2a-d). The persistence of handedness through time provides further evidence that biases are not introduced by some experimental artifact. If, for example, flies were following a wall or a trail of odors or pheromones, these results would require that they do so in a highly reproducible manner, across long time scales, and in different randomly assigned Y-mazes.

**Figure 2 – An individual's handedness is persistent over time. a-c**) Turn scores from individual flies measured in sequential experiments. Flies were assayed in the Y-mazes, stored individually, and then assayed a second time one day (**a**), two days (**b**), or four weeks (**c**) later. **d**) Correlation coefficient ($r$) of turning scores across flies tested in the Y-maze, stored individually, then tested a second time either one day, two days, one week, two weeks, or four weeks later. Error bars indicate +/- 1 standard error as estimated by bootstrap resampling. $n=85$ to $n=184$ for all time points. **d**) Mean turn bias of parental (tan) and F1 generations (brown) derived from strongly biased CS individuals (grey bars). $n$ indicates number of F1s assayed. $h^2$ indicates estimated heritability. The dashed line indicates 50%. Error bars are +/- 1 MAD, as a measure of variability rather than error. F1 and parental distributions are statistically indistinguishable.

Turning bias is evidently a persistent property of individual flies – perhaps it reflects a single source of behavioral chirality. We tested this hypothesis by measuring two additional lateralized behaviors: the direction of spontaneous exploration in circular arenas (Fig S2) and the folding arrangement of the wings at rest (Fig S3). We found that individual flies demonstrate a characteristic preference in the direction in which they circle. On average, flies spend equal amounts of time moving clockwise and counterclockwise, but individuals within the population often show strong preferences to circle in one direction or the other (Figure S2d-f). Likewise, individual flies exhibit preferences in which wing is placed on top at rest. Some fold left on top of right, others right on top of left (Fig S3a). As with the Y-mazes, both arena circling bias and wing-folding bias persist across days (Figs S2f, S3b). We tested individual flies in two assays each and found that a fly's turn bias in the Y-maze correlates significantly with a clockwise circling tendency in the arena (Fig S2g). In contrast, circling bias was completely uncorrelated to wing-folding bias. From these observations, we conclude that lateralized behavior is multifaceted, even within *Drosophila*, and that the turning biases we see in the Y-maze likely reflect an assay-independent locomotor bias phenomenon.

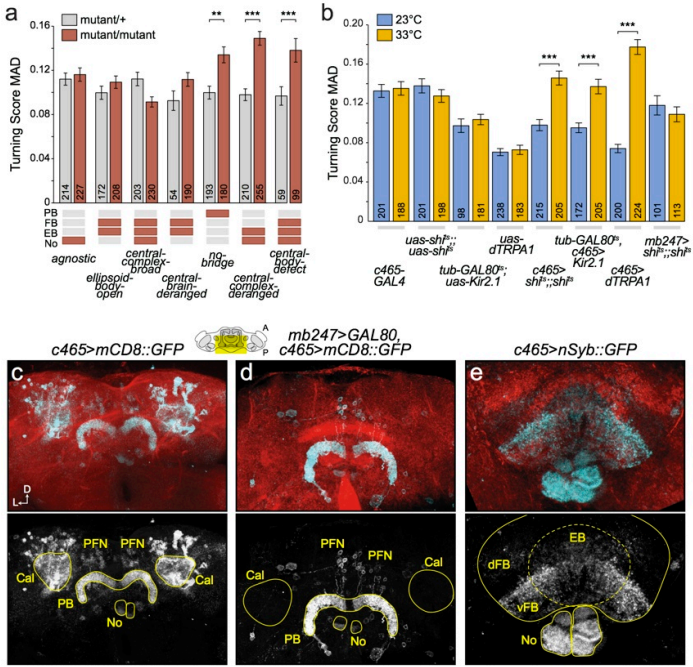

**Figure 3 – The central complex regulates variability in turn bias. a)** The degree of variability in handedness for seven fly lines carrying mutations that disrupt the development of the central complex (red bars). Three mutations significantly increase the MAD of the distribution of turning scores when compared to heterozygous controls (grey bars). ** $p<0.01$, *** $p<0.001$, as estimated by comparing bootstrap resampling of MAD values[9], Bonferroni corrected for multiple comparisons. Bars are +/- 1 standard error estimated by bootstrap resampling. Numbers indicate sample sizes. Red boxes indicate neuropils grossly disrupted by each mutation. **b)** Turn score variability (MAD) of lines with *c465-GAL4* driving expression of temperature sensitive modulators of neuronal activity (*GAL80ts;Kir2.1*, *Shibirets*, *dTRPA1*, and control lines) at 23°C (blue bars) and 33°C (orange bars) temperatures. Error bars and p-values as in **a**. **c-e)** Max fluorescence z-projections of *c465* driven expression of membrane localized (mCD8) or presynapse (nSyb) localized GFP (cyan), within the central brain (**c**) and central complex (**d**, **e**). Red counter-stain is actin. Diagram indicates anterior-posterior extent of z-projection. Cal: mushroom body calyx, PB: protocerebral bridges, No: noduli, EB: ellipsoid body, d/vFB: dorsal/ventral fan-shaped body. PFN: PB-FB-No neurons.

There are numerous possible causes of individual turning bias. One potential source of variation is the presence of polymorphic "lefty" and "righty" alleles in the population. However, locomotor handedness showed no heritability (Fig 2e, mean $h^2$=-0.03, standard error=0.018, Fisher selection test of heritability[12]), and we found no evidence that inbreeding reduces variability in locomotor bias (Figs 1f, S1e). While an individual's handedness is not heritable, the total degree of variability at the population level is under genetic control, with some lines more idiosyncratic than others (see Fig 1, Canton-S vs. $w^{1118}$ and [13]) Another potential source of persistent locomotor bias is morphological asymmetry. We examined whether variability in leg lengths could account for turning biases. We tested 28 metrics of leg length asymmetry and found that just one correlates with turning bias, and weakly ($r^2$=0.11, $p$=0.007, $p$=0.18 after multiple comparisons correction, Figure S4).

Given that neither cryptic genetic variation nor morphological asymmetry are major sources of variation in turning behavior, perhaps idiosyncratic locomotor asymmetries have a neurobiological basis. The central complex (CC) is a protocerebral structure with integral roles in processing sensory information and controlling locomotor output across arthropods[10,14-17]. We examined whether disrupting the CC can alter a population's distribution of turning biases. First, we tested seven mutants that perturb central complex development and morphology[10]. Of these, *No-bridge*, *central-complex-deranged*, and *central-body-defect* (*cbdKS96*) showed a significant increase in individual variation in turning as compared to heterozygous controls (Fig 3a). *cbdKS96* is an missense mutant of *Ten-a*[18], a transmembrane protein involved in axon targeting and synapse formation[19,20], and causes severe and widespread defects in the fan-shaped body (FB), ellipsoid body (EB), and noduli (No), leading to high individual-to-individual variation in the gross morphology of the CC[10]. Thus, variability in the structure and function of central complex circuits may give rise to variability in turn bias.

We next sought to perturb central complex function more specifically with inducible transgenes[21]. We selectively silenced subsets of CC neurons using a panel of *GAL4* lines to express a temperature inducible inhibitor of vesicle fusion (*Shibirets* or *Shits*)[22]. By comparing the median absolute deviations from the mean (MADs) of the distributions of turning scores at the permissive (23°C) and non-permissive (33°C) temperatures, we identified three *GAL4* drivers that regulate the amount of turn bias variability in a population (Figs 3b, S4-5). Acutely disrupting the function of *c465, R16D01,* or *R73D06* cells by silencing them with *Shits* caused large increases in the variability of turning scores. A similar effect resulted from acutely silencing *c465* cells with *GAL80ts;Kir2.1*[15], or by hyperactivating them with *dTRPA1*[23] (Fig 3b).

The GAL4 lines *c465, R16D01,* and *R73D06* drive expression in subsets of columnar neurons projecting from the protocerebral bridges (PB), to the fan-shaped body (FB) and contralateral No ("PFNs," Figs 3c,d, 4 and S6,7), with dendritic fields in the PB and axonal fields in the FB and No[24,25] (Fig 3e). *c465* is also expressed in the mushroom bodies, but silencing them had no effect on turn bias variability (Fig 3b). The only cell type present in all three of these lines are the PFNs (Figs 4, S6, S7). PFNs can be sub-classified into one of three types based on the regions of innervation within the FB and No[25]. Our data suggests PFNs projecting to No domain 3 may specifically be the regulators of turn bias variability (Figs 4c-d). Of the six GAL4 lines in our screen that had PFN expression, the three which had no effect all share strong expression in No domain 4 (Figs S8), hinting that silencing domain 4 PFNs might counteract or gate the effect of silencing domain 3 PFNs, a possibility which has some statistical support in our data (see Supplementary discussion).

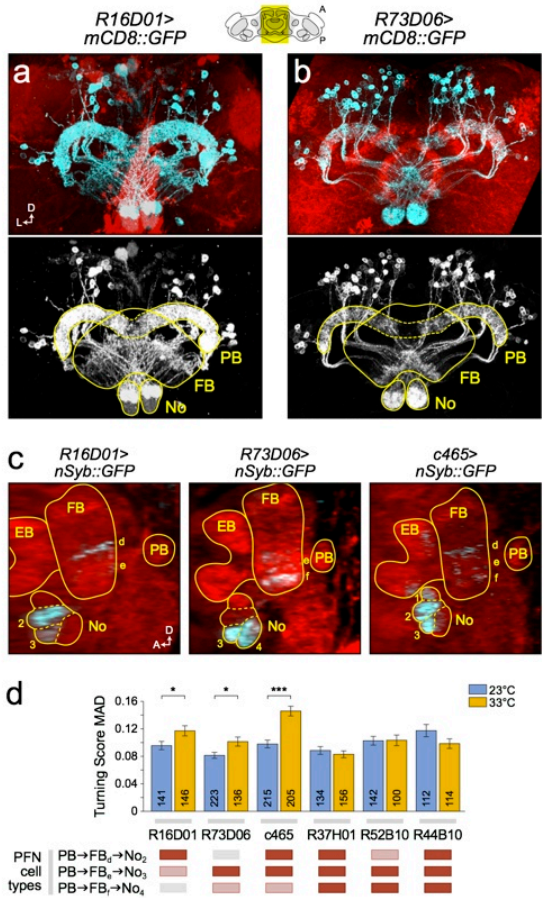

**Figure 4 – PFNs regulate variability in turn bias. a**) Max fluorescence z-projections of *R16D01-GAL4* driven expression of membrane localized mCD8::GFP (cyan), within the central complex. Red counter-stain is actin. Diagram indicates anterior-posterior extent of z-projection. **b**) As in (**a**) for the *R73D06-GAL4* driver. **c**) Lateral views of CD8::GFP driven by *R16D01*, *R73D06* and *c465-GAL4*. PB: protocerebral bridges, FB: fan-shaped body, No: noduli, EB: ellipsoid body. d, e, f: layers of the ventral fan-shaped body, 1, 2, 3, 4: domains of the noduli. **d**) Turn score variability of lines with various GAL4 lines driving shibire at 23°C (blue bars) and 33°C (orange bars). Bars are +/- 1 standard error estimated by bootstrap resampling. Numbers indicate sample sizes. * $p<0.05$, *** $p<0.001$. Red boxes indicate PFN subtypes with high GAL4 expression, pink boxes lower expression.

Our results suggest that genetically and environmentally matched fruit flies exhibit individual differences in the neural processing of sensory information and the execution of locomotor patterns, resulting in profound levels of idiosyncratic handedness. Specifically, columnar PFN neurons of the central complex may be involved in the integration of bilateral sensory information, or the modulation of stimulus signal-to-noise ratios, and individual asymmetries in their functions may result in asymmetric behavioral outputs (see Supplementary discussion and Fig S9). I.e., when a fly must make a left vs. right decision in the absence of an asymmetric stimulus, asymmetries within the brain predispose the animal to go one way rather than the other. Perhaps this can help the animal avoid analysis paralysis, or perhaps it is a feature of noisy biological systems. Individual variation in wiring[26-28], physiology[29] and behavior[9,30] may prove to be a very general feature of neural circuits, with broad implications both for our basic understanding of developmental neurobiology and the emergence of behavioral phenotypes at the individual level.

**Methods**

All raw data, data acquisition software and analysis scripts are available at *http://lab.debivort.org/neuronal-control-of-locomotor-handedness/*.

Fly lines – Flies were housed on modified Cal Tech medium (either from KD Medical or the Harvard University BioLabs fly food facility) according to standard protocols. A full list of the lines used in this study is available in the Supplementary Information. Flies used for *Shibire*[ts] experiments were reared at 25°C and transferred to 33°C 30 minutes prior to and during data collection. *GAL80*[ts];*Kir2.1* experimental groups were reared at 18°C, transferred to 30°C for 48 hours prior to testing, and transferred to 33°C for data collection. Controls were kept at 18°C until testing at 33°C.

Maze fabrication – Mazes were cut into 1/16" thick black acrylic using a laser engraver (Epilog). Each arm of a maze was 0.37" long and 0.13" wide. In order to inhibit the flies from flipping upside down, the floors of the mazes were lightly roughened with a random orbital sander and fine-grit sand paper, and clear acrylic lids (one per maze) were lubricated with Sigmacote (Sigma). Circular arenas were fabricated similarly and were 2 inches in diameter. A diffuser made of two sheets of 1/4" thick clear acrylic roughened on both sides by orbital sanding, placed between the LS array and the maze array, provided for uniform illumination.

Behavioral tracking and data analysis – Flies were placed into individual Y-mazes or arenas and allowed to walk freely for 2hrs. Mazes were illuminated from below with white LSs (5500K, LuminousFilm), imaged with 2MP digital cameras (Logitech, Point Grey), and the X-Y positions of the flies' centroids were automatically tracked using background subtraction and recorded with software custom written in LabView (National Instruments). Data were then analyzed with custom written scripts in MatLab (The MathWorks). Fly identity was maintained for day-to-day experiments by storing flies individually in labeled culture vials between tests.

Wing-folding – To measure the wing-folding preference of individual flies, we moved each fly into a vial singly. The vial was flicked or agitated until the fly flew, assuring that its wings were unfolded. We then anesthetized it with CO2, and examined it manually under a dissecting scope to determine which wing was on top. The animal was then returned to its vial and allowed to waken, at which point we repeated the process. Flies were examined in this way 5 times sequentially per day. To generate the day-to-day correlation (Fig S3b), we compared the aggregated wing-folding data from days 1 and 2 (10 total observations) with the aggregated data from days 3 and 4 (another 10 total observations). Similarly, the flies that were first measured for wing-folding and then turning bias in the Y-maze were scored for the former over two days (10 total observations) and tested in the Y-maze on day 3.

Statistics – Expected distributions (Figs 1e, S1e) were calculated by summing binomial distributions with $n_i$ equal to the number of choices made by fly $i$ within the corresponding experimental group, and $p_i$ equal to the average right-turn probability of the entire population. These individual curves were interpolated into a normalized [0,1] domain before summing. Standard errors of MAD scores and correlation coefficients were calculated using bootstrap resampling, with a minimum of 1000 replicates. P-values reported based on bootstrap resampling in three different ways: 1) To compare observed MADs to known null hypotheses (Fig 1e,f), samples were drawn from the known null distribution in numbers corresponding to the data. The number of samples in which the MAD of the randomly drawn values equal or was less than that of the null hypothesis was recorded. If the bootstrap replicates produced the tested condition by chance alone at a rate of $k$ out of $n$ resamples, $p$ was reported as the highest probability such that $\sum_{j=0}^{k} \binom{n}{j} p^j (1-p)^{n-j} > 0.025$. That is, the highest value of $p$ that would yield the tested event $k$ times out of $n$ or fewer at least 2.5% of the time. 2) To compute bootstrapped z-tests (e.g. Fig S4b), we determined the number of standard errors away from 0 the observed MAD was by bootstrapping the points contributing to that MAD value. 3) To compare the MADs of two experimental groups (Figs 3a-b 4d, and S6) we assumed the bootstrap estimated errors on the MADs were Gaussian and calculated the 1-tailed probabilities of that a MAD drawn from the experimental error distribution would be less than a MAD drawn from the control error distribution. Mutual information (Fig S1f) was calculated on a fly-by-fly basis between turn $t$ and $t+1$. Significance asterisks in Fig 3a reflect a Bonferroni correction for multiple comparisons. Significance in Figs 3b and 4d is not corrected since comparisons are only between 23°C and 33°C experimental groups.

Immunohistochemistry and imaging – Adult nervous tissue was dissected and fixed overnight in 4% paraformaldehyde at 4°C. After fixation, tissue was counterstained with Alexa Fluor 568 conjugated to phalloidin for 24-48hrs (Life Technologies, 1:50 dilution). Stained brains were washed in PBT, and mounted on glass slides in 70% glycerol or vectaShield mounting medium (Vector Labs). Images were collected on a Zeiss LSM710 or LSM780 confocal microscope. Panels modified from FlyLight images (Figs S7d,e and f right side, and S8 bottom row) were downloaded from *http://flweb.janelia.org/cgi-bin/flew.cgi* and color rotated into a red-cyan palette. Depth coded images (Fig S7f) and lateral views (Figs 4c and S8) were calculated using stack functions in FIJI (*http://fiji.sc/Fiji*).

Modeling – Model simulations (Fig S9) were performed in MATLAB (The MathWorks) using Euler approximation, with $\Delta t$=0.01. The model was considered converged (i.e. a "decision" had been made) if variables changed by less than 0.001 between iterations, or at 1000 iterations, whichever was earlier. Tuning curves were determined empirically based on 1000 replicates of the model. Behavioral distributions were based on 1000 tuning curves. Beta fitting parameters were determined analytically from the means and variances of the behavioral distributions. Additional details are given in Extended Experimental Procedures. The model was highly robust to parameter choice, and the parameter values used for Figure 6 are $\alpha_L=\alpha_R=0.01$, $\beta_L=\beta_R=0.02$, $\delta=0.03$, $\gamma_R=0.01$, and $\gamma_L=0.01b$, where $b$ determines the intrinsic network bias. Stimulus noise was implemented as $L'_{in}(0)=L_{in}(0)+\varepsilon_1$ and $L'_{in}(0)=L_{in}(0)+\varepsilon_2$ where $\varepsilon_i \sim N(\mu,\sigma^2)$, with $\mu$=0 in all cases, $\sigma$=0.1 for the green curves, and $\sigma$=0.2 for the purple curves of Figure 6. Intrinsic bias was normally distributed with mean=1 and standard deviation deviation=0.025 in Figure S9c and 0.01 in S9d and e.

**Conflict of interest:**

The authors have no financial interests related to this work.

**Acknowledgements:**

This research was funded by the Junior Fellows Program at The Rowland Institute at Harvard. We would like to thank Mike Burns, Chris Stokes, and other members of The Rowland Institute for fruitful scientific discussions and technical assistance, as well as Frank Hirth, Tom Maniatis, Charles Zuker, and members of their labs for helpful feedback. We thank Shmuel Raz, Roland Strauss, Michael Reiser, Aravi Samuel, Sam Kunes, Chuntao Dan, Douglas Armstrong and Hiromu Tanimoto for sharing fly lines. We thank the Janelia Farm FlyLight consortium for allowing us to reuse and modify their *GAL4* expression images.

**Supplementary materials**

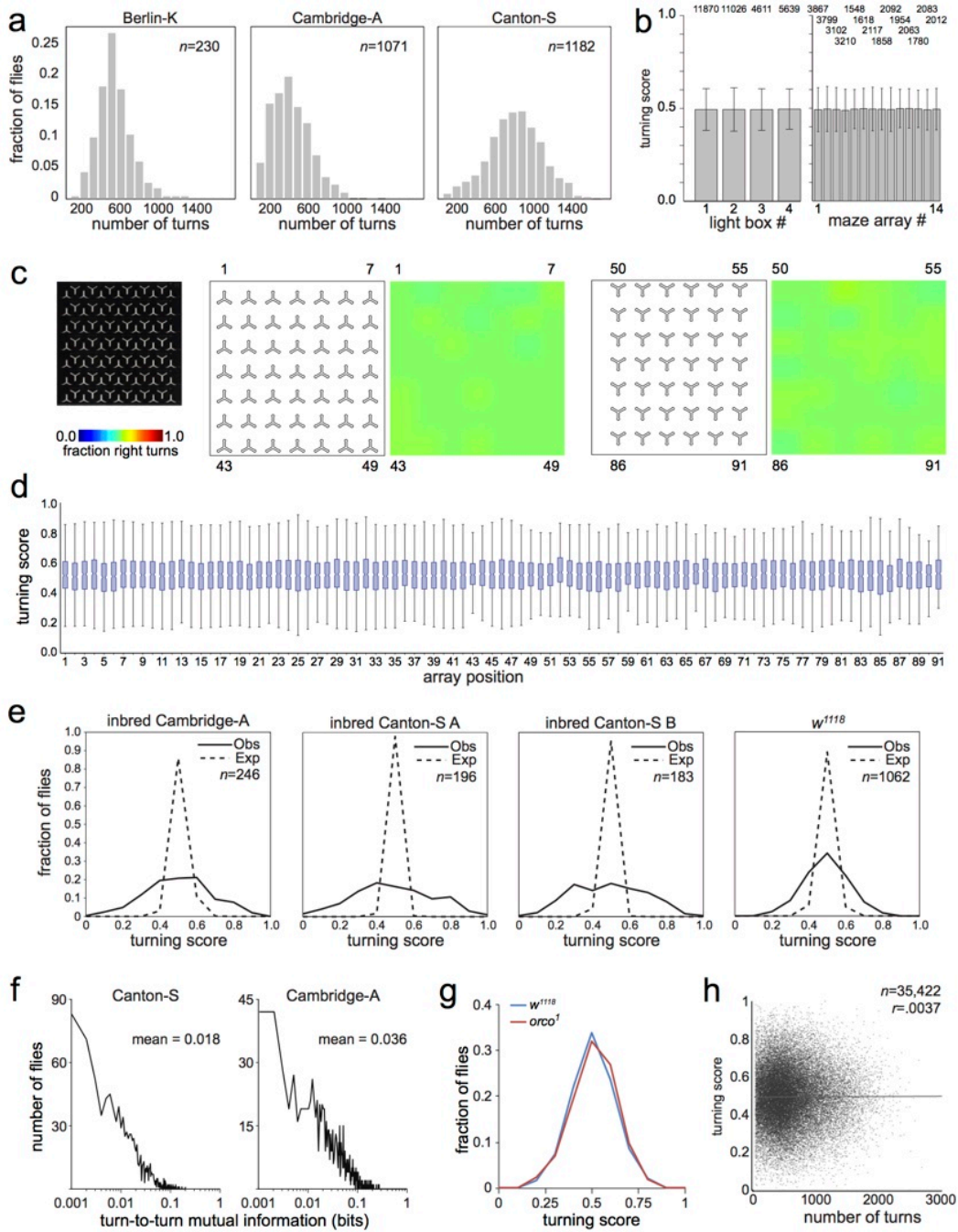

**Supplementary Figure 1 – Y-maze assay performance and controls. a**) The distribution of the total number of turns completed in 2 hours for three wild type lines. **b**) Experiments were carried out with 14 different maze arrays containing grids of either 49 or 91 individual mazes, and imaged on 4 light box and camera imaging rigs. Left) Average turning score +/- MAD for all flies assayed in each of the 4 light box and camera setups. Numbers indicate sample sizes. Right) Average turning score +/- MAD for all flies assayed in each of the 14 arrays of Y-mazes. **c**) Example photo of an array of 91 mazes (left). Average turning score for all flies in each of the first 49 maze array positions (center-left diagram and heat map), and for mazes in positions 50-91 (right diagram and heat map). Green in heat maps indicates turn biases of 0.5. **d**) 1-way ANOVA analysis of all flies run at each position in the maze arrays. No position has a significantly different distribution of turning scores, $p=0.450$. Waist of box plot indicates median turning score. Box edges indicate 25th and 75th percentiles. Whiskers indicate range excluding outliers. **e**) The distribution of observed (solid lines) and expected (dotted lines) turning scores for four WT fly lines. **f**) Fly-by-fly distributions of the mutual information between successive turns in the Y-maze, for two wild type lines. Mutual information equal to 1 denotes complete dependence of each turn on the previous turn, 0 complete independence. **g**) Olfactory input does not alter the distribution of turning scores. The distribution of turning scores was determined for a broad spectrum olfactory mutant ($orco^1$, $n=212$) and the corresponding background line ($w^{1118}$, $n=1062$) **h**) The number of completed turns does not correlate to turning score.

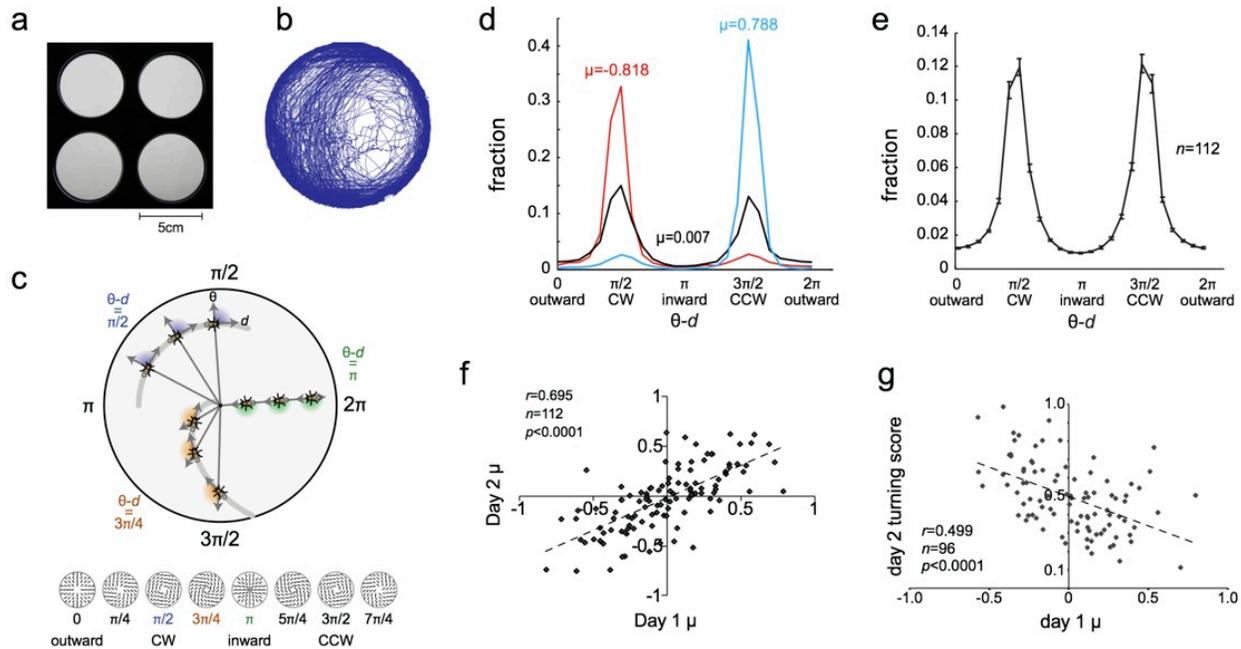

**Supplementary Figure 2 – Individual flies exhibit circling bias in an open arena. a)** An open arena assay for exploratory locomotion. Individual flies were placed in circular arenas, allowed to walk freely for 2 hours, and their position was tracked as in Figure 1. **b)** Example path data collected from a single fly over 2 hours. **c)** For each data point, a circling score is calculated by subtracting the fly's direction of motion (d) from its angular position (θ) in radians. This gives the circumferential component of motion, with π/2 indicating clockwise motion (CW), 3π/2 indicating counterclockwise motion (CCW), 0 indicating walking straight into the center of the arena, and π indicating walking straight out from the center. **d-e)** Histograms of circling scores for a strongly CW-biased individual, a CCW-biased individual, and a relatively unbiased individual (**d**), and the average circling of 112 flies (**e**). **f)** Measures of average circling are correlated between testing days across flies. μ is the averaged signed circumferential component of motion (vel$_{circum}$/speed); 1 corresponds to purely CCW motion, -1 corresponds to purely CW motion and 0 corresponds to equal (or no) CW and CCW motion. **g)** Circling scores (μ) as measured in the arena are correlated to turning bias as measured in the Y-maze. Flies were assayed in the open arenas, stored individually, then assayed 24 hours later in the Y-mazes.

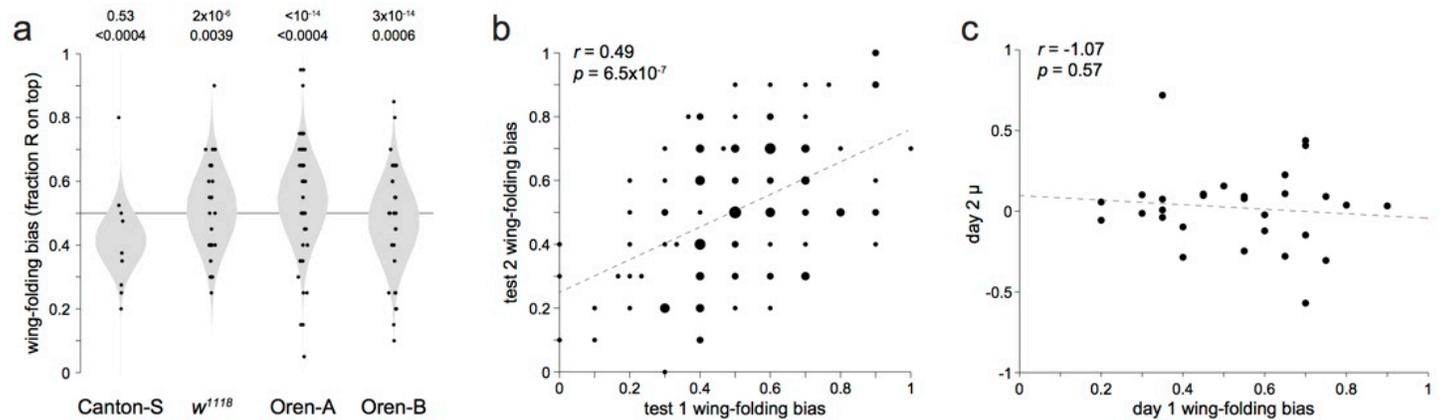

**Supplementary Figure 3 – Individual flies exhibit wing-folding bias. a)** Wing-folding bias scores (points) for individual flies. Grey distributions reflect null hypotheses based on the number of times each fly was assessed and the mean observed across all flies. Top row of *p*-values are from the $\chi$2 test of variance, bottom row from bootstrap resampling. **b)** Bubble plot of wing-folding scores over subsequent days of testing (days 3 and 4 vs. 1 and 2, see Methods). The areas of points reflect the number of superimposed data points. Data is aggregated from the four lines shown in (**a**); the same positive trend was present in all lines individually. **c)** Scatter plot of arena circling score (μ) versus wing-folding score for individual flies measured in both assays and whose identities were preserved by single housing.

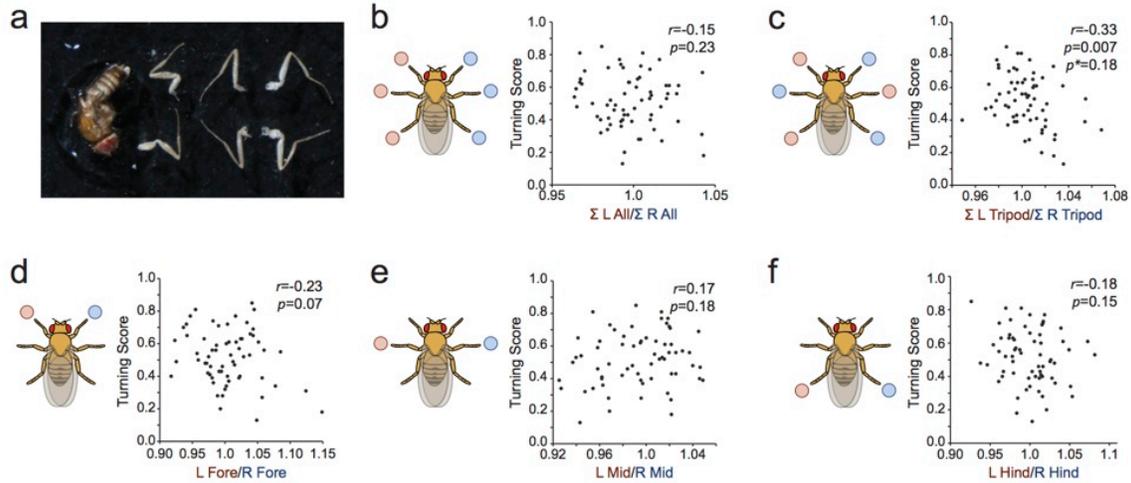

**Supplementary Figure 4 – The contribution of leg morphology to handedness. a**) 65 flies were assayed in the Y maze and preserved in 100% ethanol. Legs were removed, photographed, and the lengths of the leg femurs, tibia and tarsi measured, along with the length of the body. Shown is a representative preparation for measurement. **b-f**) Scatter plots where each point represents an individual's turning score versus the ratio of (**b**) the total length of left legs vs. the total length of right legs, (**c**) the total length of the left tripod vs. the right tripod, (**d**) the left foreleg vs. the right foreleg, (**e**) the left midleg vs. the right midleg, (**f**) the left hindleg vs. the right hindleg. 23 other correlations between turning score and individual leg or segment length were also considered, so the Bonferroni correction for multiple comparisons in (**c**), *p\**, reflects that number of comparisons.

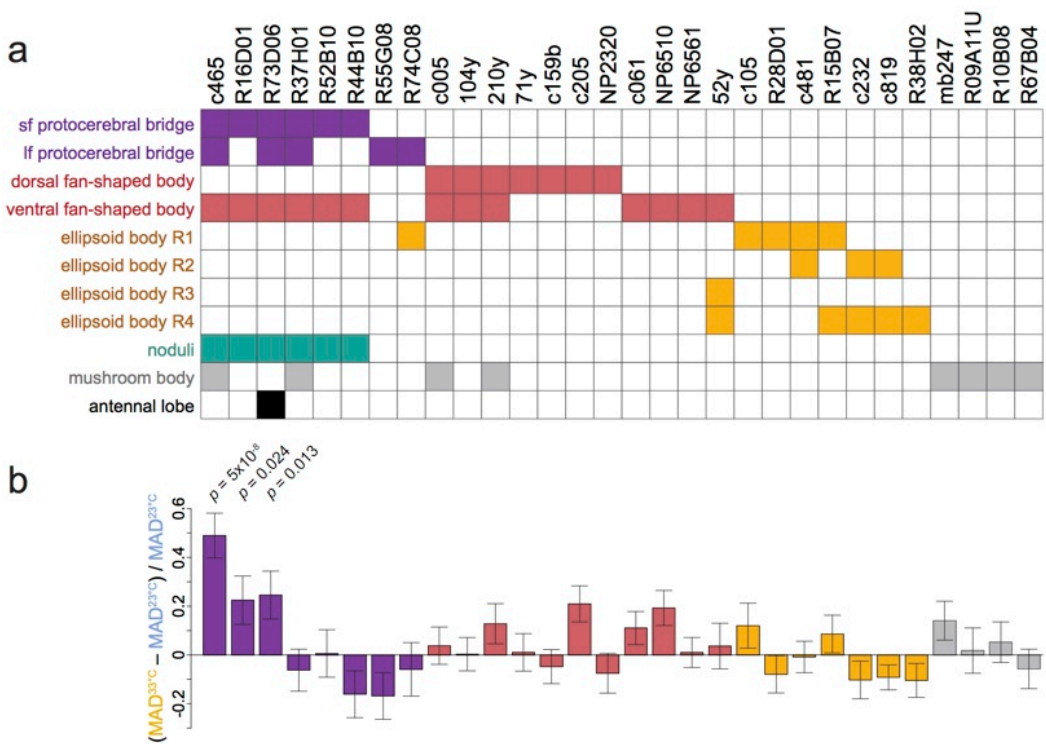

**Supplementary Figure 5 – A screen of *GAL4* lines for neurons affecting handedness**. **a-b**) 30 different sparsely expressed *GAL4* lines were used to drive temperature sensitive S*hibire* in subsets of central complex (and other brain regions). Colored cells in table (**a**) indicate regions of expression. (**b**) For each line, % difference in the MAD of turning scores between the restrictive (33°C) and permissive (23°C) temperatures. Error bars indicate error estimated by bootstrap resampling. *P*-values for the three lines with effects (top rows) were determined by a *z*-test on the number of standard errors between the estimate of the % MAD difference and 0.

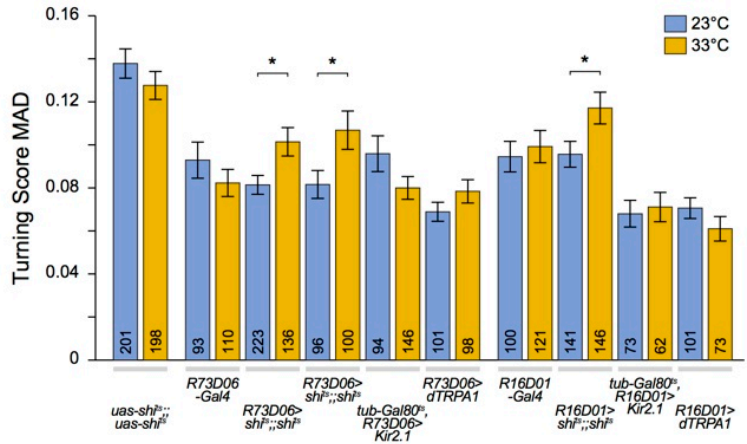

**Supplementary Figure 6 – PFN GAL4 lines regulating handedness**. Turn score variability (MAD) of lines with either *R73D06* or *R16D01-GAL4* driving expression of temperature sensitive modulators of neuronal activity (*GAL80$^{ts}$;Kir2,1*, *Shibire$^{ts}$*, *dTRPA1*, and control lines) at 23°C (blue bars) and 33°C (orange bars) temperatures. Unlike *c465-GAL4* these drivers had no effect when driving *Kir2.1* or *dTRPA1*, their baseline effect sizes were also lower. Column pairs 3 and 4 represent two separate *R73D06>Shibire$^{ts}$* conducted a few weeks apart. Bars are +/- 1 standard error estimated by bootstrap resampling. Numbers indicate sample sizes.

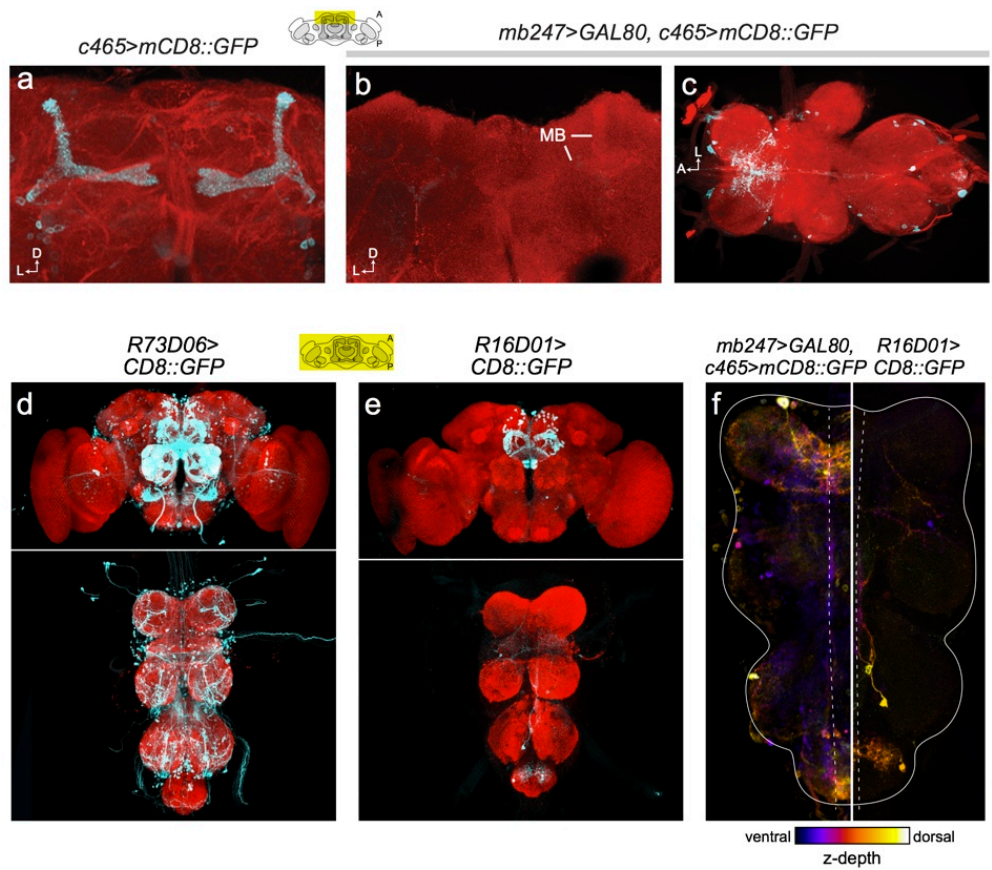

**Supplementary Figure 7 – GAL4 expression pattern analysis**. **a**) Max fluorescence z-projection of *c465* driven expression of CD8::GFP (cyan), within the central brain. Red counter-stain is actin. Diagram indicates anterior-posterior extent of z-projection in (**a**) and (**b**). Projection as in (**a**) in a fly with *c465-GAL4* and *MB247-GAL80* driving CD8::GFP expression. All staining in the mushroom bodies (MB) is eliminated. **c**) Max projection through the entire A-P axis of the ventral nerve cord (VNC), same genotype as (**b**). **d**) Max projections through the entire brain (top) and VNC (bottom) of CD8::GFP driven by *R73D06-GAL4*. **e**) As in (**d**) for the driver *R16D01-GAL4*. **f**) Depth-coded image of CD8::GFP expression as driven by *c465-GAL4;MB247-GAL80* (left) and *R16D01* (right). There are no cells common to both expression patterns within the VNC. Images in (**d**), (**e**), and (**f**, right) reproduced and modified with permission from published FlyLight translation stacks *http://flweb.janelia.org/cgi-bin/flew.cgi*.

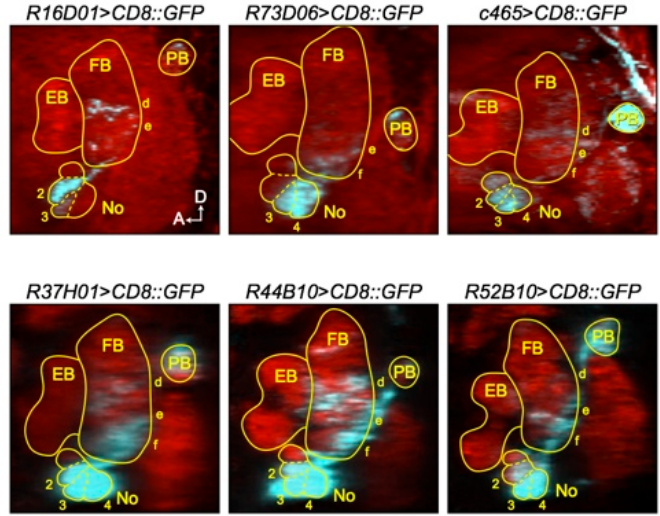

**Supplementary Figure 8 – PFN subtype analysis**. Single slice lateral views through the central complex reconstructed from confocal stacks. On the medial-lateral axis, slices are centered on the noduli. CD8::GFP driven by labeled *GAL4* drivers (cyan). Red counter-stain in the top row is actin and anti-nc82 staining in the bottom row. PB: protocerebral bridges, FB: fan-shaped body, No: noduli, EB: ellipsoid body. d, e, f: layers of the ventral fan-shaped body, 1, 2, 3, 4: domains of the noduli. Images in the bottom row reproduced and modified with permission published from FlyLight translation stacks *http://flweb.janelia.org/cgi-bin/flew.cgi*.

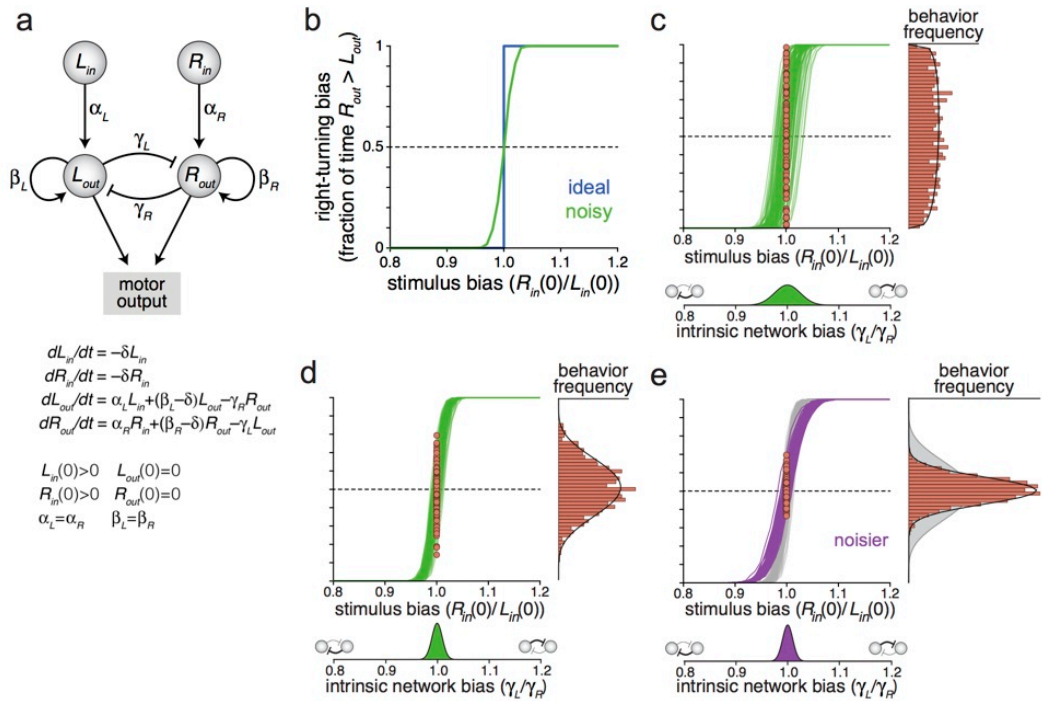

**Supplementary Figure 9 – A model of idiosyncrasy in locomotor decision making. a**) Mutual-inhibition model of a decision making circuit. Activity in $L_{in}$ and $R_{in}$ represents the summed stimuli favoring left and right turning, respectively. These nodes activate $L_{out}$ and $R_{out}$ (with strengths $\alpha_L$ and $\alpha_R$) which are reciprocally inhibitory with strengths $\gamma_L$ and $\gamma_R$. After the model state has converged, motor behavioral output is scored as "left" if $L_{out} > R_{out}$ and vice-versa. Terms of the model, initial conditions and assumptions are given. **b**) The model generates "ideal" decisions in the absence of noise (i.e. any small bias in stimulus results in a complete motor bias). Introducing Gaussian noise to the stimulus results in a more plausible sigmoidal stimulus-locomotor tuning curve. See Methods and Supplementary Discussion for details. **c**) Introducing individual-to-individual variation in the "intrinsic bias" of the network (*e.g.* by skewing the ratio of reciprocal inhibitory strengths $\gamma_L$ and $\gamma_R$) results in individuals exhibiting a variety of left-right biases under identical stimulus circumstances (red points, which are binned into a histogram at right). Y-axes as in **b**; fit (black line) is provided by a beta function. **d**) Reducing individual-to-individual network asymmetry reduces behavioral variability as compared to (**c**). **e**) Decreasing the stimulus signal-to-noise ratio flattens the stimulus-locomotor tuning curve, and reduces variability in behavioral outcomes. Grey data are from (**d**) and shown for comparison.

**Movie S1 – A Y-maze locomotor choice assay.** A movie demonstrating a single fly exploring a Y-maze for 2 minutes. As the fly passes through the center of the maze (green box) its centroid is tracked (red circle) and its path is recorded (blue lines). Passes through the choice point are scored as left or right turns in post processing. *https://www.youtube.com/watch?v=RtRmiP0lN7U*

**Movie S2 – An array of Y-mazes in the imaging rig.** A movie demonstrating 91 individual flies simultaneously exploring separate Y-mazes. Each array of 91 mazes was constructed out of laser cut acrylic, evenly illuminated from below with diffuse white light, and imaged from above. *https://www.youtube.com/watch?v=VZdVwpnVwpI*

| Strain | Source |
|---|---|
| Berlin-K | BSC #8522 |
| Cambridge-A | lab stocks[1] |
| Cambridge-A iso | lab stocks[1] |
| Canton-S | W. Quinn (MIT) |
| Canton-S isoA | this work |
| Canton-S isoB | this work |
| white[1118] | J. Dubnau (CSHL) |
| Oren-A | S. Raz (IEUH) |
| Oren-B | S. Raz (IEUH) |
| agnostic[X1] | R. Strauss (JGUM) |
| central-body-defect[KS96] | R. Strauss (JGUM) |
| central-brain-deranged[849] | R. Strauss (JGUM) |
| central-complex-broad[KS145] | R. Strauss (JGUM) |
| central-complex-deranged[KS135] | R. Strauss (JGUM) |
| ellipsoid-body-open[678] | R. Strauss (JGUM) |
| no-bridge[KS49] | R. Strauss (JGUM) |
| norpA-/- | A. Samuel (HU) |
| orco[1] | BSC #23129 |
| UAS-dTRPA1 | BSC #26263 |
| UAS-GFP.nls | BSC #4775 |
| tubP-Gal80[ts];UAS-Kir2.1 | M. Reiser (JFRC) |
| UAS-mCD8-GFP | BSC #5136 |
| UAS-nSyb-GFP | S. Kunes (HU) |
| UAS-Shibire[ts];;UAS-Shibire[ts] | C. Dan (JFRC) |
| 52y-GAL4 | D. Armstrong (UE) |
| 71y-GAL4 | D. Armstrong (UE) |
| 104y-GAL4 | D. Armstrong (UE) |
| 210y-GAL4 | D. Armstrong (UE) |
| c005-GAL4 | D. Armstrong (UE) |
| c061-GAL4 | D. Armstrong (UE) |
| c105-GAL4 | D. Armstrong (UE) |
| c159b-GAL4 | D. Armstrong (UE) |
| c205-GAL4 | D. Armstrong (UE) |
| c232-GAL4 | D. Armstrong (UE) |
| c465-GAL4 | D. Armstrong (UE) |
| c481-GAL4 | D. Armstrong (UE) |
| c819-GAL4 | D. Armstrong (UE) |
| mb247-GAL4 | S. Kunes (HU) |
| mb247-GAL80 | H. Tanimoto (MPIN) |
| NP2320-GAL4 | DGRC #104157 |
| NP6510-GAL4 | DGRC #113956 |
| NP6561-GAL4 | DGRC #105258 |
| R09A11U-GAL4 | M. Reiser (JFRC) |
| R10B08-GAL4 | M. Reiser (JFRC) |
| R15B07-GAL4 | M. Reiser (JFRC) |
| R28D01-GAL4 | M. Reiser (JFRC) |
| R38H02-GAL4 | M. Reiser (JFRC) |
| R67B04-GAL4 | M. Reiser (JFRC) |

**Table S1 – Strains used in this study and their origins.** BSC: Bloomington Stock Center; CSHL: Cold Spring Harbor Laboratory; DGRC: Drosophila Genetic Resource Center; HU: Harvard University; JFRC: Janelia Farm Research Campus; JGUM: Johannes-Gutenberg Universitaet Mainz; IEUH: Institute of Evolution, University of Haifa; MIT: Massachusetts Institute of Technology; MPIN: Max Planck Institute for Neurobiology; UE: University of Edinburgh.

**Supplementary discussion**

Statistical support for differential effects of PFN subtypes – Three *GAL4* lines with expression in the PFNs modify turn bias variability when perturbed transgenically (*c465*, *R73D06* and *R16D01*). Our screen (Fig S5) also contained three lines (*R37H01*, *R52B10*, *R44B10*) with expression in PFNs which had no effect (Fig 4d). While these lines may have had no effect for all the usual experimental reasons – e.g. not driving *Shibire* strongly enough or not being expressed in a sufficient number of PFNs – we saw some indications that the lines that had no effect shared aspects of their expression patterns (Fig 4d). Specifically, these lines all had strong expression in domains 4 (and 3) of the noduli. Perhaps the effect of silencing PFNs is dose-dependent, so that silencing an intermediate number of PFNs, or all PFNs at an intermediate level, results in modulation of turn bias variability, whereas silencing them all or very strongly undoes the effect.

Alternatively, there may be a gating or additive relationship between the PFN subtypes, such that silencing PFNs projecting to domain 3 of the noduli increases turn bias variability, but simultaneously silencing PFNs projecting to domain 4 reduces variance, or blocks the effect of perturbing domain 3 PFNs. We are limited in our ability to resolve these hypotheses by a lack of *GAL4* lines that express uniquely in domain 3 or domain 4. To our knowledge the six *GAL4* lines we examined are all the available drivers for targeting PFNs specifically. The development of more specific genetic tools will allow us to experimentally test hypotheses pertaining to the respective contributions of individual neuronal subtypes.

In the meantime, even though the number of independent *GAL4* lines is low, we did conduct a simple statistical test of the idea that silencing domain 3 PFNs increases variability while silencing domain 4 PFNs reduces it. Specifically, we used the expression coding from Fig 4d (bottom), and fit a linear model where the % change in MAD from permissive to restrictive temperatures of a particular *GAL4* line = $a$*D2 + $b$*D3 + $c$*D4, where *a*, *b* and *c* were fit based on the observed data, and D2, D3 and D4 represent the relative level of expression in that GAL4 line in noduli domains 2, 3 and 4 respectively, coded as 0 if there was no detectable expression, 1 for weak or intermediate expression, and 2 for strong expression. This was done across 10,000 bootstrap resamplings in which the individual flies going into each experimental group were resampled with replacement, and which *GAL4* lines (out of the six) were used to fit *a*, *b* and *c*, were also resampled with replacement. Any resamples in which three or fewer *GAL4* lines were chosen were overdetermined and rejected.

Across these bootstrap replicates, 95% confidence interval values of *a*, *b*, and *c* were estimated respectively as (-0.51, 0.28), (0, 1.29), (-1.0, 0) with means -0.052, 0.51 and -0.49. The mean $r^2$ on the linear fit was 0.58 and $p = 0.047$. Thus, there is some statistical support for the notion that silencing PFNs projecting to domain 4 may reduce turn bias variability, while silencing PFNs projecting to domain 3 increases variability. Definitive tests of this hypothesis will be possible when new genetic reagents allow us to target specific PFN subtypes.

A model of asymmetry and noise in left-right decision making – We were curious how a symmetrical perturbation (e.g. silencing the PFNs) could enhance both left- and right-biased asymmetries. To investigate this we explored a general decision making circuit model, and found that circuit perturbations that decrease the signal-to-noise-ratio of the stimuli triggering choice behavior result in an increase in behavioral variability by increasing the relative importance of any small inherent network asymmetry. We found no other perturbation that had this effect. Thus, a reasonable hypothesis for the effects we observed is that silencing the PFNs increases the signal-to-noise ratio in the stimuli driving locomotor choice behavior. Our purpose here was not to reinvent known models of reciprocal inhibition, but rather to rigorously test our intuition that a plausible explanation of the effects of silencing PFNs is through a modulation of stimulus signal-to-noise ratios.

It is striking that perturbation of these neurons results in the exacerbation of asymmetric locomotor behaviors in a way that affects left- and right-bias equivalently. The overall symmetry of the distribution of turning scores is maintained, suggesting that any minor asymmetries in this expression pattern are incidental to its symmetric effect on the frequency of left and right-biased animals. Likewise, it is unlikely that the effect of silencing the PFNs is due to consistent asymmetry in their structure (i.e. like the asymmetric body[2]), as silencing or hyperactivating an asymmetrical circuit in a background of symmetrical behaviors would likely introduce asymmetry detectable in the average behavior.

To explore circuit dynamics that might underlie these observations, we developed an average firing rate model of a simple left-right decision making circuit (Fig S9). This model consists of two input units whose activity reflects the aggregated stimuli in favor of left or right turning respectively. Each input activates one of two outputs that inhibit each other and activate themselves. The activity of these output units determines the motor output – if the left output unit is active, then a left motor instruction is generated and vice-versa. Reciprocal inhibition circuits similar to this one have been implicated in sensory signal processing in vertebrates and invertebrates (e.g.,[3,4]).

Over a wide range of parameters, this circuit amplifies any difference between left and right stimuli, and produces a pattern of all-or-none left vs. right output (Fig S9b). This is, of course, biologically implausible. Even in strongly stimulus-evoked behaviors, animals exhibit a variety of behavioral outcomes. This trial-to-trial stochasticity can be built into the model by adding a random error term to the initial stimulus values (Fig S9b). This reshapes the stimulus-locomotor tuning curve from an idealized square wave to a more plausible sigmoidal shape. That is, when the input stimuli are strongly biased, the output behavior approaches 100% left or 100% right turns, but when stimuli are ambiguous, arbitrary left-right behavioral probabilities can be attained.

In our Y-maze experiments, animals were presented with essentially identical, highly ambivalent stimuli. There were no systematic cues to drive the animals left or right. Nevertheless, flies exhibited a broad distribution of left-right turning biases. In the framework of our model, this observation can be recreated by introducing processing asymmetries across individuals (Fig S9c-e). In one implementation, we varied according to a normal distribution the ratio of the strengths of the reciprocal inhibitory connections between the output units. This resulted in a population of tuning curves, and consequently a broad distribution of left-right behavioral outcomes induced by ambiguous stimuli. Tuning the variance in processing asymmetry has a direct effect on the breadth of the behavioral distribution (which in all cases was well fit by a beta function, Fig S9d,e). This corresponds to the entire population of flies with its diverse collection of turning biases. Thus, the amount of locomotor variation specific to each strain can be accounted for by a particular amount of individual-to-individual variation in neural substrates, either in their physiology or wiring.

With the circuit model now reflecting levels of behavioral variability specific to each line, we can investigate manipulations that broaden or narrow the behavioral distribution while holding constant the magnitude of inter-animal network asymmetries. Many symmetrical manipulations, such as proportionally scaling the strengths of the reciprocal inhibitory connections (or self-activating connections, or activating connections from input to output units) have no effect. However, modulating the input signal-to-noise ratio, by manipulating the error term in the stimulus encoding, profoundly affects the breadth of the behavioral distribution (Fig S9e). Interestingly, decreasing the signal-to-noise ratio narrows the behavioral distribution and flattens the population of sigmoidal tuning curves as they pass through the regime of ambivalent stimuli. In other words, decreasing the salience of whatever stimuli induce the animals to turn either left or right causes their choices to be based that much more on the random error in the encoding of those stimuli. Across many trials, this will drive behavior toward its mean across animals, narrowing the distribution of behaviors.

Our model assumes that a network of reciprocal inhibitory connections amplify differences in input stimuli, resulting in a winner-takes-all output. Such an inhibitory network may be present in the ellipsoid body which harbors reciprocally inhibitory connections between sister ring neurons (Ludlow et al., 2014, manuscript submitted). Upstream neuropils of the EB within the CC include the FB and the PB, implicating the c465 neurons as inputs to the decision-making circuit. The model specifically suggests that perturbing the activity of the c465 neurons results in an increase in the signal-to-noise ratio of the stimuli driving turning behavior. Silencing these neurons (with *Shibire$^{ts}$*), or saturating their activity to the point of effective uniformity (with *dTRPA1*) would increase signal-to-noise if c465 neurons were contributing to the noise component of that ratio. Activity in the protocerebral bridges may encode information useful for turning decision-making in circumstances other than in the experimental setting of our Y-mazes. Indeed, visual information flows via at least two routes to the central complex: from the polarized light-sensitive dorsal rim ommatidia to the protocerebral bridges[5], and from the higher-order feature detectors of the optic lobe and optic glomeruli to the lateral triangle (and EB)[6]. Since there are no polarized light sources in our experiment, it is plausible that activity in c465 neurons constitutes noise with respect to useful visual stimuli (Figure 7G). Since presynapses of c465 neurons are found in the ventral FB and No, these neuropils may be the sites at which stimuli relevant and irrelevant for locomotor turn decision-making are integrated.